\documentclass[aoas,MSNbibl,nameyear,dvips]{arximspdf}

\doi{10.1214/14-AOAS728ED} 
\referstodoi{10.1214/14-AOAS728}
\volume{8}
\issue{4}
\pubyear{2014}
\firstpage{1921}
\lastpage{1921}
\docsubty{FLA}


\begin{document}
\begin{frontmatter}

\title{Editorial: Spatial accessibility of pediatric primary healthcare: Measurement and inference}
\runtitle{Editorial}
\pdftitle{Editorial: Spatial accessibility of pediatric primary healthcare: Measurement and inference}

\begin{aug}
\author[A]{\fnms{Susan M.}~\snm{Paddock}\corref{}\ead[label=e1]{paddock@rand.org}}
\runauthor{S. M. Paddock}
\affiliation{RAND Corporation}
\address[A]{RAND Corporation\\
1776 Main Street\\
Santa Monica, California 90401\\
USA\\
\printead{e1}} 
\end{aug}

\received{\smonth{9} \syear{2014}}


\end{frontmatter}

Improving access to health care has long been at the forefront of policy
debates in the U.S. There are multiple determinants of healthcare
utilization: predisposing characteristics that explain individuals'
propensities to use healthcare; enabling characteristics that describe the
resources individuals have to use healthcare; and perceived or actual need
for healthcare [\citet{AdaAnd74}]. \citet{NobSerSwa14}
illustrate the complexity involved with developing an understanding of one
determinant of healthcare utilization. They examine spatial accessibility---an
enabling characteristic under the aforementioned framework---to
primary care pediatricians in Georgia. The authors encounter challenges
that arise in many public policy applications, namely, the limitations of
the available data, the need to conduct analyses that reflect system
constraints, model selection and uncertainty quantification. The Area
Editors featured this paper, along with contributions from three
discussants, in an AoAS invited session at the 2014 Joint Statistical
Meetings and are including the paper and those discussions in this issue
because the paper showcases the type of research AoAS aims to publish:
analyses requiring innovative statistical thinking to address questions of
practical importance.



\printaddresses
\end{document}